\documentclass[preprint,showpacs,preprintnumbers,amsmath,amssymb,floatfix]{revtex4-1}
\usepackage{color}
\usepackage{graphicx}
\usepackage{dcolumn}
\usepackage{bm}
\usepackage{multirow}

\definecolor{sienna}{cmyk}{0,0.72,1,0.45}
\definecolor{fg}{cmyk}{0.91,0,0.88,.12}
\definecolor{yellow}{cmyk}{0,0,1,0}
\definecolor{or}{cmyk}{0,1,0.5,0}
\definecolor{magenta}{cmyk}{0,1,0,0}
\definecolor{rubinered}{cmyk}{0,1,0.13,0.45}
\definecolor{blue}{cmyk}{1,1,0,0}
\definecolor{turquoise}{cmyk}{1,1,0,0.5}
\definecolor{aquamarine}{cmyk}{0,1,0,0.0}
\definecolor{midnightblue}{cmyk}{1,0.5,0.0,0.0}
\definecolor{junglegreen}{cmyk}{1,0,0.2,0.5}

\begin{document}

\title{The effect of distributed time-delays on the synchronization of neuronal networks}

\author{Ajay Deep Kachhvah}
\affiliation{Indian Institute of Science Education and Research (IISER) Mohali, Knowledge City, Sector 81, SAS Nagar,  Manauli PO 140 306, Punjab, India.}

\begin{abstract}
Here we investigate the synchronization of networks of FitzHugh-Nagumo neurons coupled in scale-free, small-world and random topologies, in the presence of distributed time delays in the coupling of neurons. We explore how the synchronization transition is affected when the time delays in the interactions between pairs of interacting neurons are non-uniform. We find that the presence of distributed time-delays does {\em not} change the behavior of the synchronization transition significantly, vis-a-vis networks with constant time-delay, where the value of the constant time-delay is the mean of the distributed delays. We also notice that a normal distribution of delays gives rise to a transition at marginally lower coupling strengths, vis-a-vis uniformly distributed delays. These trends hold across classes of networks and for varying standard deviations of the delay distribution, indicating the generality of these results. So we conclude that distributed delays, which may be typically expected in real-world situations, do not have a notable effect on synchronization. This allows results obtained with constant delays to remain relevant even in the case of randomly distributed delays.

\end{abstract}

\maketitle

\section{Introduction}

\noindent The phenomenon of synchronization is one of the most important example of self-organized coordination between individual dynamical units in many realistic and man-made complex systems. This phenomenon provides insights for understanding collective dynamical behavior in many different physical and biological systems such as the flashing of fireflies, the rhythmic pacemaker cells of the heart, respiration, power grids, social phenomena \citep{kuramoto,strogatz,glass,pikovsky,arenas}. 
In particular, synchronization is very crucial for the transmission of information through network of coupled neurons across the nervous system. So exploring synchronization of the activity of model neuronal oscillators may assist in providing valuable insights into the understanding of the dynamics of the brain. 

Now, in a physical system when an information signal from one location reaches another through a transmission line, there is always a time-delay in the received signal. Namely, a time-delay in coupling arises due to inherent finite propagation signal speeds \cite{choi,yeung,ares,herrgen}. Such time-delayed coupling in different complex networks has been studied quite extensively \citep{perez,peron,eguiluz,wang,wang1,wang2}, with most of the research being focussed on the effect of constant time-delay or time-varying delay between agents \citep{cao,hunt,hunt2,kohar,sugitani,wille,selivanov}.

In this work our objective is to study the {\em effect of randomly distributed time-delays on the synchronization process in complex networks with different connection topologies}. Our central question is the following: how is the synchronization transition in a network affected when the time-delay between pairs of connected nodes in the network is not same, but rather is distributed randomly. It is also relevant to ask if the type of distribution (for instance, Gaussian vis-a-vis uniform) is significant in determining the collective behavior. 

Thus our aim here is to explore the synchronization process on various network topologies with randomly distributed time-delays, and to ascertain how the emergent behavior under distributed delays is different from the behavior of the same network under constant time-delay. In particular, as a test-bed of our investigation, we consider networks of neurons, modeled by the well-known FitzHugh-Nagumo (FHN) system. In our system then, the FHN neurons are coupled in different classes of network topology, i.e. the neuronal dynamics at the nodes of the network is given by the FHN model and the adjacency matrix of the network, representing the connectivity of the neurons, ranges from random to scale-free and small-world networks. Each pair of connected neurons has an information time-delay whose value is drawn from a distribution such as a uniform or normal (Gaussian) distribution. We will demonstrate through extensive simulations below that the synchronization transition is {\em not significantly affected by the random distribution of delays, in all classes of network topologies and types of delay distributions}.

\section{Network of Model Neurons}

\noindent We consider the FitzHugh-Nagumo model neuron \citep{fitz,nagumo} at the nodes of the network. The FHN model provides the simplest representation of firing neuronal dynamics and has been widely used as a model for the spiking neurons \citep{eugene}. The spatiotemporal evolution of a network of such model neurons, with information time-delay, is governed by the equations \citep{buric,murray}:
\begin{equation}
\dot{v_i}(t) = v_i(t)[v_i(t)-a][1-v_i(t)] - w_i(t) + I + K\sum_{j=1}^N A_{ij} [v_j(t-\tau_{ij}) - v_i(t)],
\label{fhn_fast}
\end{equation}
\begin{equation}
\dot{w_i}(t) = \epsilon[v_i(t) - b w_i(t)].
\label{fhn_slow}
\end{equation}
where $i=1...N$ denotes neurons in the network of size $N$, $v_i(t)$ is the membrane potential of the $i$th neuron and  $w_i(t)$ is the variation of its ion concentration. These two variables, $v$ and $w$, represent the fast and the slow variables of the neuron model, with parameter $\epsilon$ being small enough to give rise to the slow temporal evolution of $w_i(t)$. Parameter $K$ gives the coupling strength between neurons, and parameter $a$ crucially determines the dynamics of the individual neurons.

We study three different interaction networks: (i) random network, proposed by E{\H r}dos-R{\'e}nyi \cite{renyi}; (ii) scale-free network, proposed by Barab{\'a}si-Albert \cite{barabasi} and (iii) small-world network, proposed by Watts-Strogatz \cite{watts}. So the dynamics at the nodes of the network is governed by FitzHugh-Nagumo equations Eq.(\ref{fhn_fast}), where the adjacency matrix element $A_{ij}$ is $1$ if neuron $i$ is connected to neurons $j$, and $A_{ij}=0$ otherwise. The nature of the adjacency matrix is naturally determined by the class of network being considered. 

The important network parameters of the different network classes are as follows. For random networks parameter $P$ gives the probability for link creation and determines the number of links, and thus the over-all connectivity, of the system. For small-world networks the most relevant parameter is $p$, the probability of rewiring each link which determines the fraction of random links in the network. For the scale-free network, an important network characteristic is the exponent $\lambda$ in the power-law degree distribution. 

The $\tau_{ij}$ is the information time-delay between neuron $i$ and neuron $j$. For every realization, we have different values of time-delay at each link in the network, with the values $\tau_{ij}$ being drawn from a specified distribution, for instance the normal (Gaussian) or the uniform distribution.

In order to study synchronization transition, we consider a synchronization order parameter $\sigma$ \cite{gao}. This is given in terms of the time-averaged standard deviation of the fast variable $v_i(t)$, and is specifically:
\begin{equation}
\sigma = \frac{1}{T}\sum_{t=1}^T\sigma(t), \sigma(t)=\frac{1}{N}\sum_{i=1}^N[v_i(t)]^2 - \Bigg[\frac{1}{N} \sum_{i=1}^N v_i(t)\Bigg]^2.
\label{sync_para}
\end{equation}

This quantity $\sigma$ is an excellent measure of spatiotemporal synchronization in extended systems. From Eq.(\ref{sync_para}) it is evident that the more synchronous the neural network, the smaller the synchronization parameter $\sigma$. Accordingly, in the event of complete synchrony we have $\sigma=0$, and incoherence leads to large $\sigma$.

\section{Synchronization Transition}

\noindent Now we present the results from extensive numerical simulations, for random, small-world, and scale-free networks of FHN neurons, with distributed time-delays $\tau_{ij}$. In all simulations, the values of the parameters are $a=0.139$, $b=2.54$, $\epsilon=0.001$, and the external current $I=0.03$, unless mentioned otherwise. The values of the information time-delay $\tau_{ij}$ are drawn from either uniform or normal distributions, in the range $0$ to $20$. The synchronization order parameter $\sigma$ is averaged over $20$ independent runs for each set of parameter values, for statistical accuracy.

\begin{figure}[htb]
\begin{center}
\begin{tabular}{cccccc}
(a)&
\hspace{-0.8cm}
\includegraphics[height=5cm,width=5.5cm]{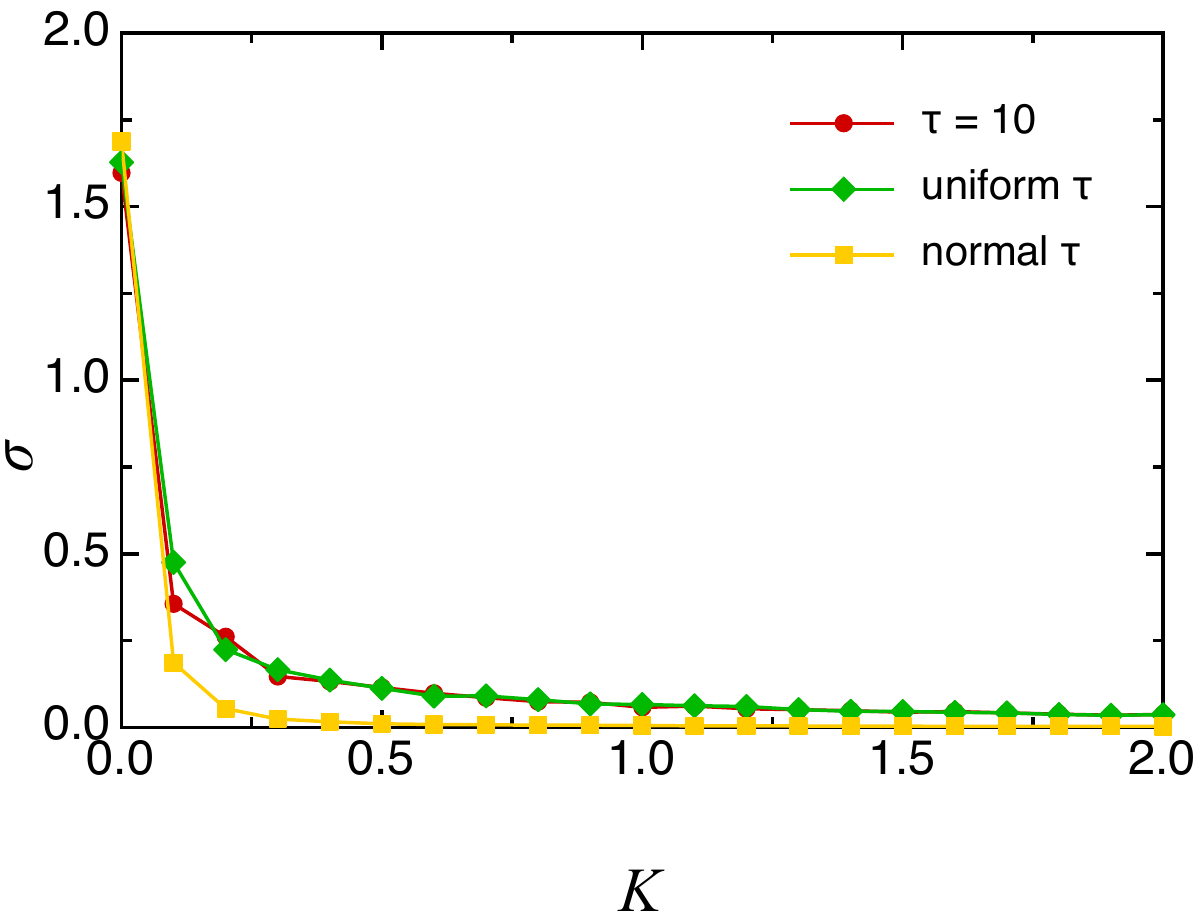}
(b)&
\hspace{-0.8cm}
\includegraphics[height=5cm,width=5.5cm]{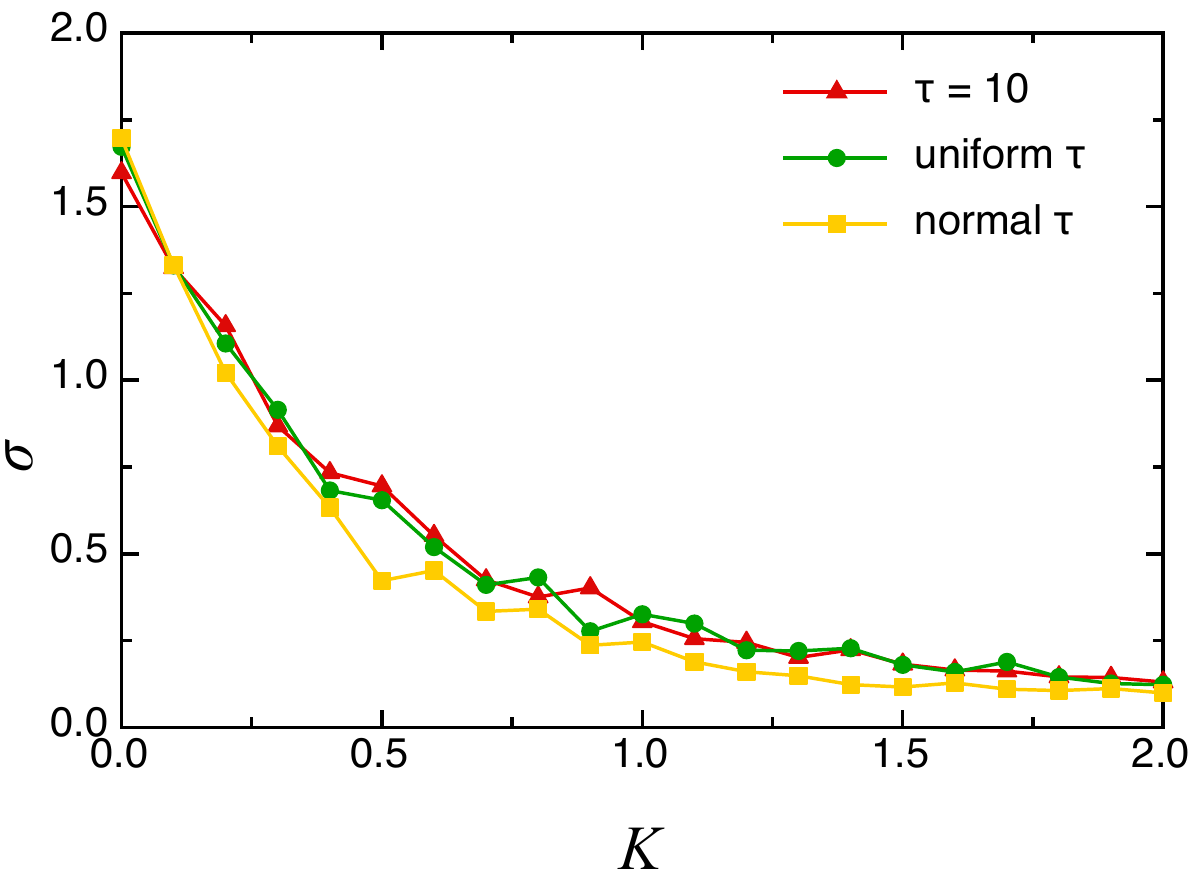}&
(c)&
\hspace{-0.8cm}
\includegraphics[height=5cm,width=5.5cm]{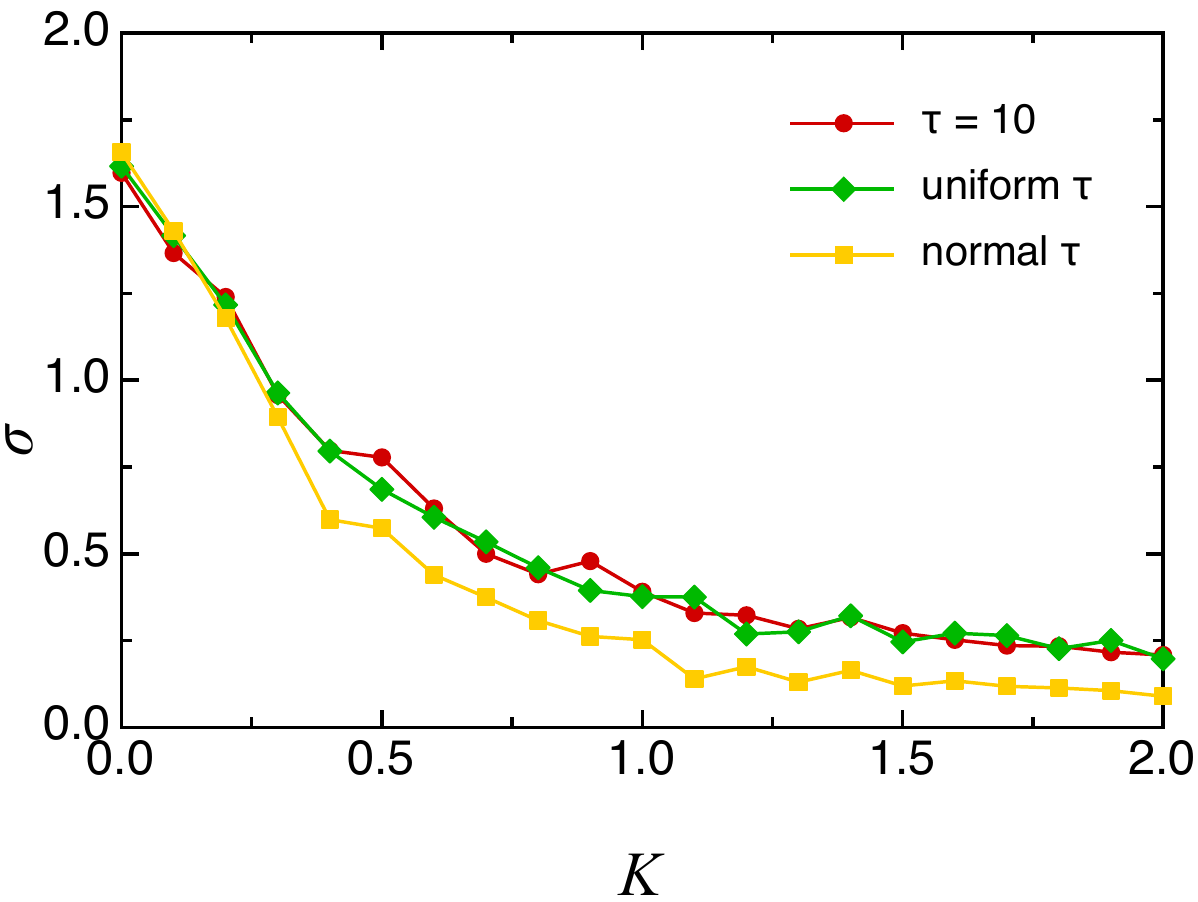}&
\end{tabular}{}
\caption{ Synchronization order parameter $\sigma$ with respect to the coupling strength $K$, for (a) random network with $P=0.4$, (b) scale-free network with $\lambda=3$, and (c) small-world network with $p=0.03$, when the information time-delay $\tau_{ij}$ is fixed at constant value $10$, drawn from a uniform distribution centered at $10$ and from a normal distribution centered at $10$ with variance $5$. Here system size $N=100$.}
\label{fig:fhn_sp_tau10}
\end{center}
\end{figure}

In order to investigate the synchronization transitions that arise for the case of distributed information time-delays $\tau_{ij}$s in the network of FHN neurons, we compute the synchronization parameter $\sigma$ defined in Eq.\ref{sync_para} as a function of the coupling strength $K$. Fig.\ref{fig:fhn_sp_tau10} presents a comparison of the synchronization transitions arising in the three different network topologies: (a) random (ER), (b) scale-free (BA) and (c) small-world (WS) networks. In each panel three curves are given, displaying the three cases of interest: (i) when all values of the time-delays are fixed at a constant value, namely all $\tau_{ij}=10$, (ii) when $\tau_{ij}$s are randomly distributed, with the random values being drawn from a uniform distribution and (iii) when $\tau_{ij}$s are randomly distributed, with the random values being drawn from a normal distribution. Note that the distributions have mean value equal to the value of the constant delay in (a), namely the random distributions are centered around $\tau_{ij} = 10$. From the panels in the figure it is clearly evident that, for all the three classes of networks, there is {\em no significant change in the nature of the synchronization transition} when the $\tau_{ij}$s are randomly distributed as compared to constant time-delays. The only perceptible difference is a marginal shift in the synchronization curves towards lower coupling strengths for the case of normal distribution, as compared to the cases fixed delays and uniformly distributed delays.

\begin{figure}[htb]
\begin{center}
\hspace{-0.3cm}
\includegraphics[height=6cm,width=7.5cm]{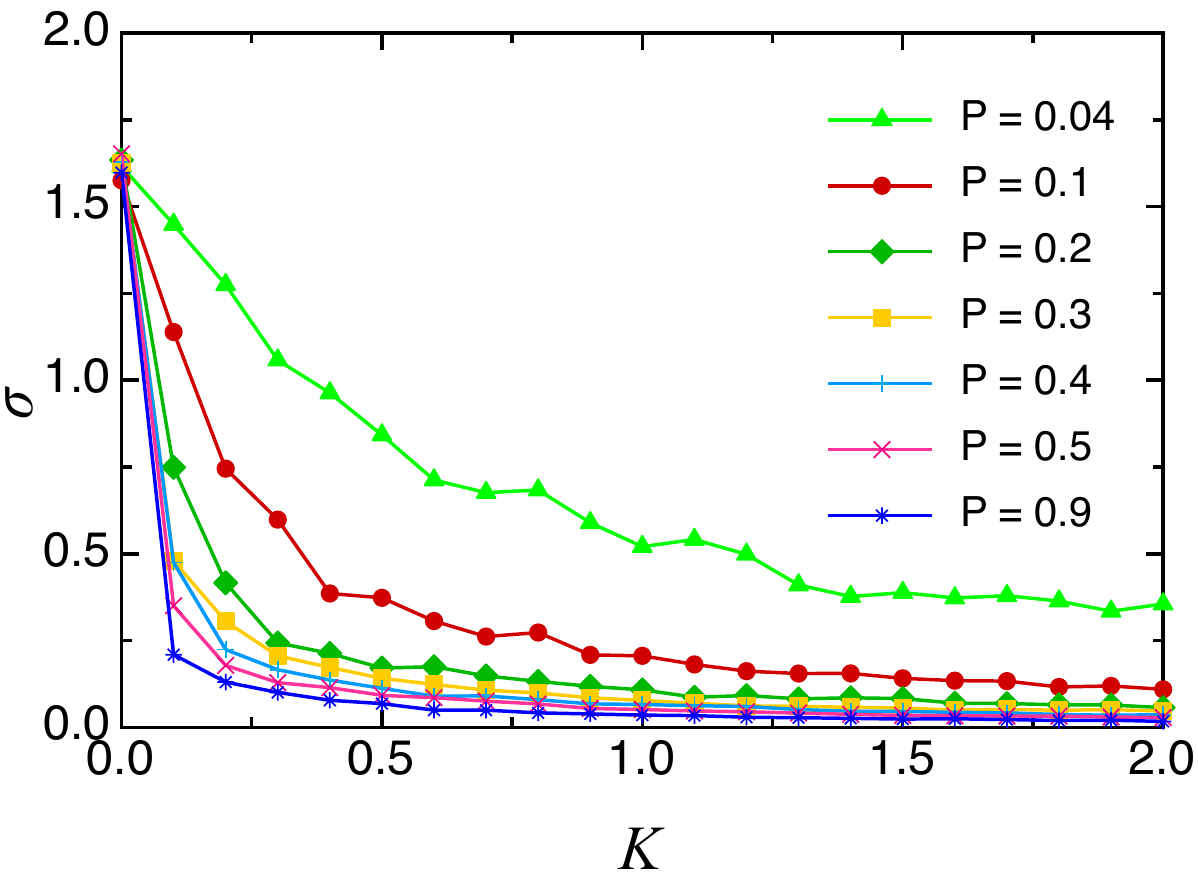}
\caption{ Synchronization parameter $\sigma$ as a function of coupling strength $K$ for E{\H r}dos-R{\'e}nyi random networks of $N=100$ neurons generated with different values of connection probability $P$. The information time-delays $\tau_{ij}$ are drawn from a uniform random distribution.}
\label{fig:fhn_er_p_sp}
\end{center}
\end{figure}

{\em Synchronization Transition in Random Networks:} Now we focus specifically on how synchronization transitions in E{\H r}dos-R{\'e}nyi random networks with distributed delays are affected by the probability for link creation $P$. Fig.\ref{fig:fhn_er_p_sp} exhibits the effect of the probability $P$ in random networks on the synchronization order parameter $\sigma$. We observe that as $P$ increases, the system synchronizes more efficiently, with the effect saturating after a high enough $P$. This is as expected, as increasing $P$ results in an increase in the number of connections, and increasing links aids synchronization. Again the distribution of delays does not have any significant influence of the synchronization features, other than a marginal lowering of $\sigma$ in networks with normally distributed delays.

\begin{figure}[htb]
\begin{center}
\begin{tabular}{cccc}
(a)&
\hspace{-0.3cm}
\includegraphics[height=6cm,width=7.5cm]{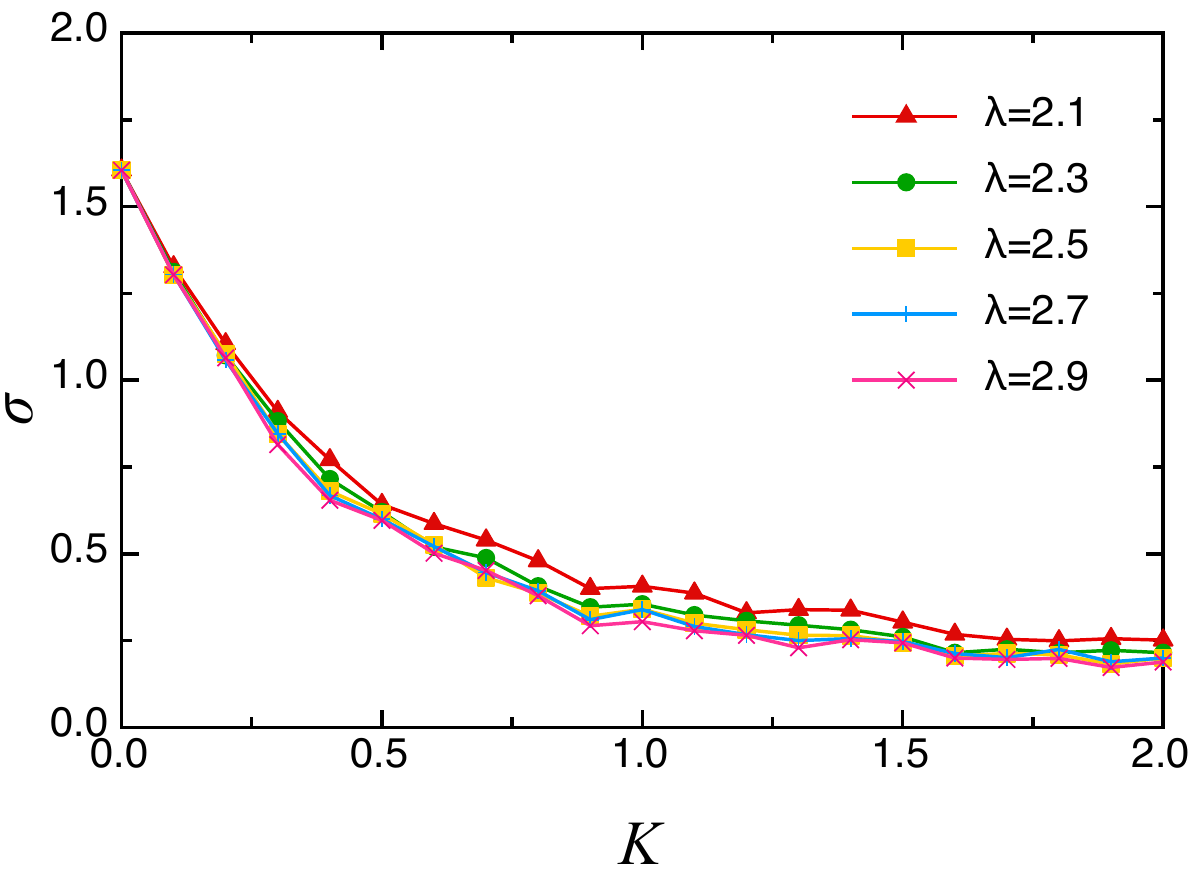}&
(b)&
\hspace{-0.3cm}
\includegraphics[height=6cm,width=7.5cm]{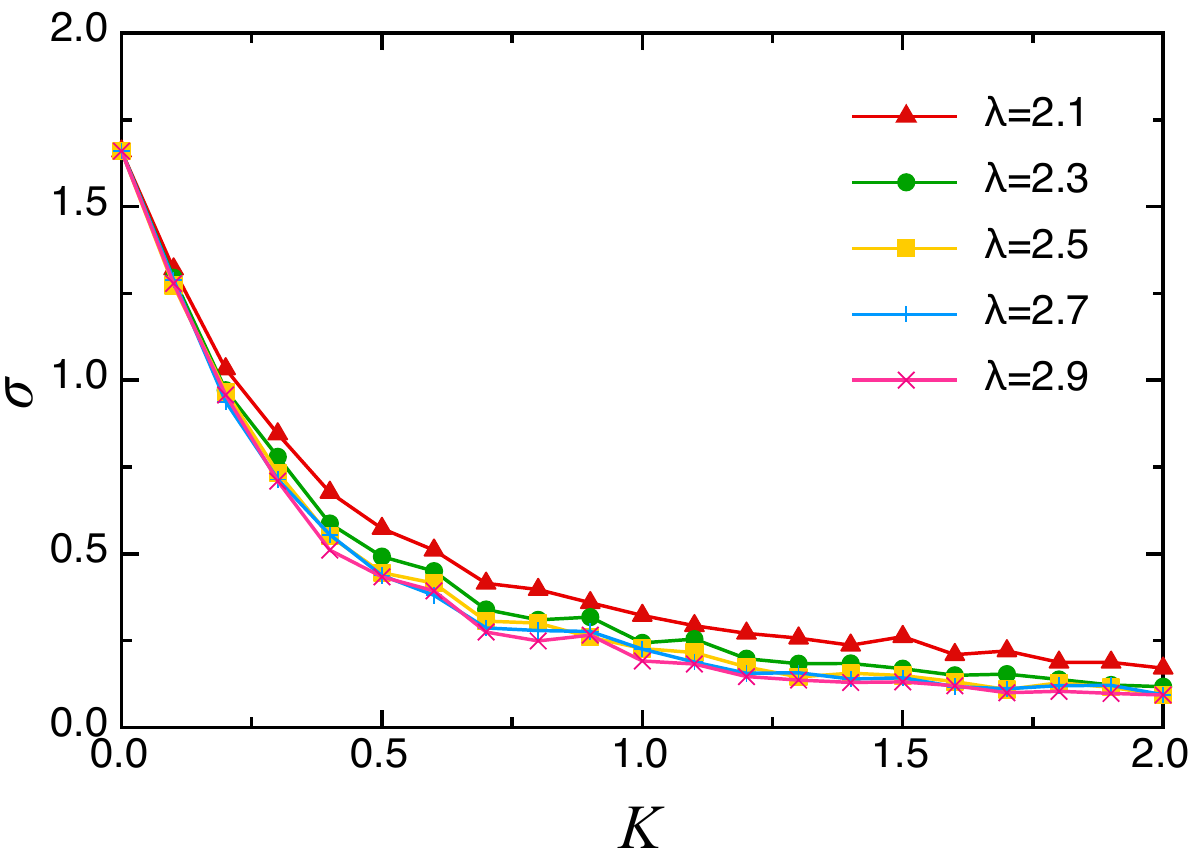}
\end{tabular}{}
\caption{Synchronization order parameter $\sigma$ with respect to the coupling strength $K$, for scale-free networks (generated using the configuration model) with power law exponents $\lambda=2.1, 2.3, 2.5, 2.7$ and $2.9$, when the information time-delay $\tau_{ij}$ is drawn randomly from (a) a uniform distribution, and  (b) a normal distribution with variance equal to $5$. The mean of both distributions is the same. Here system size $N=100$.}
\label{fig:fhn_sf_conf}
\end{center}
\end{figure}

{\em Synchronization Transition in Scale-Free Networks:} Now we probe the synchronization transition in scale-free networks with different values of the power law exponent $\lambda$ of its degree distribution. Such networks are  generated using configuration model \citep{goh,cho}, not the BA algorithm. It can be observed from Fig.\ref{fig:fhn_sf_conf} that, as the power law exponent $\lambda$ increases the synchronization parameter $\sigma$ decreases, i.e., when the degree distribution of the scale-free network falls more sharply, synchronization occurs more efficiently. One can again see from the figure that normal distribution of delays allows the network to synchronize at a slightly lower coupling strength than constant delays or a uniform distribution of delays.

\begin{figure}[htb]
\begin{center}
\begin{tabular}{cccc}
(a)&
\hspace{-0.3cm}
\includegraphics[height=6cm,width=7.5cm]{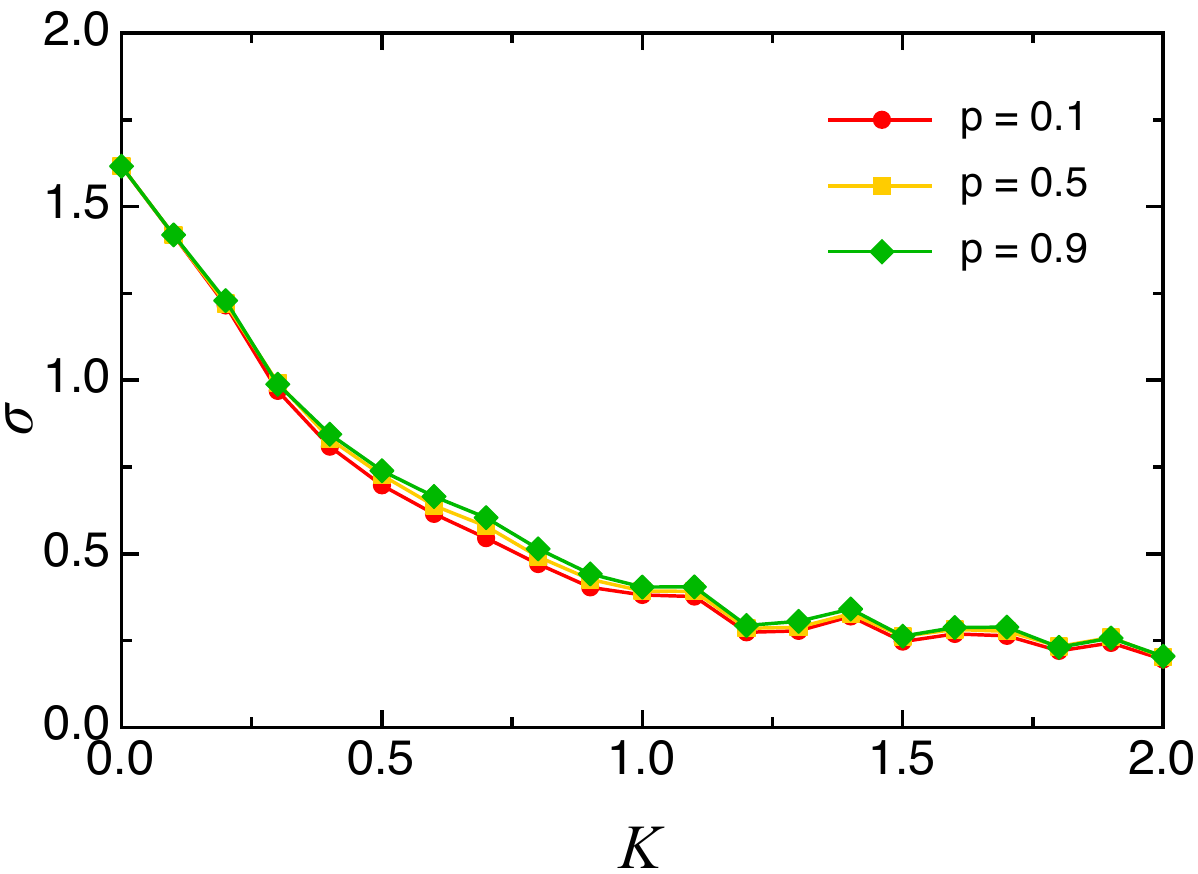}&
(b)&
\hspace{-0.3cm}
\includegraphics[height=6cm,width=7.5cm]{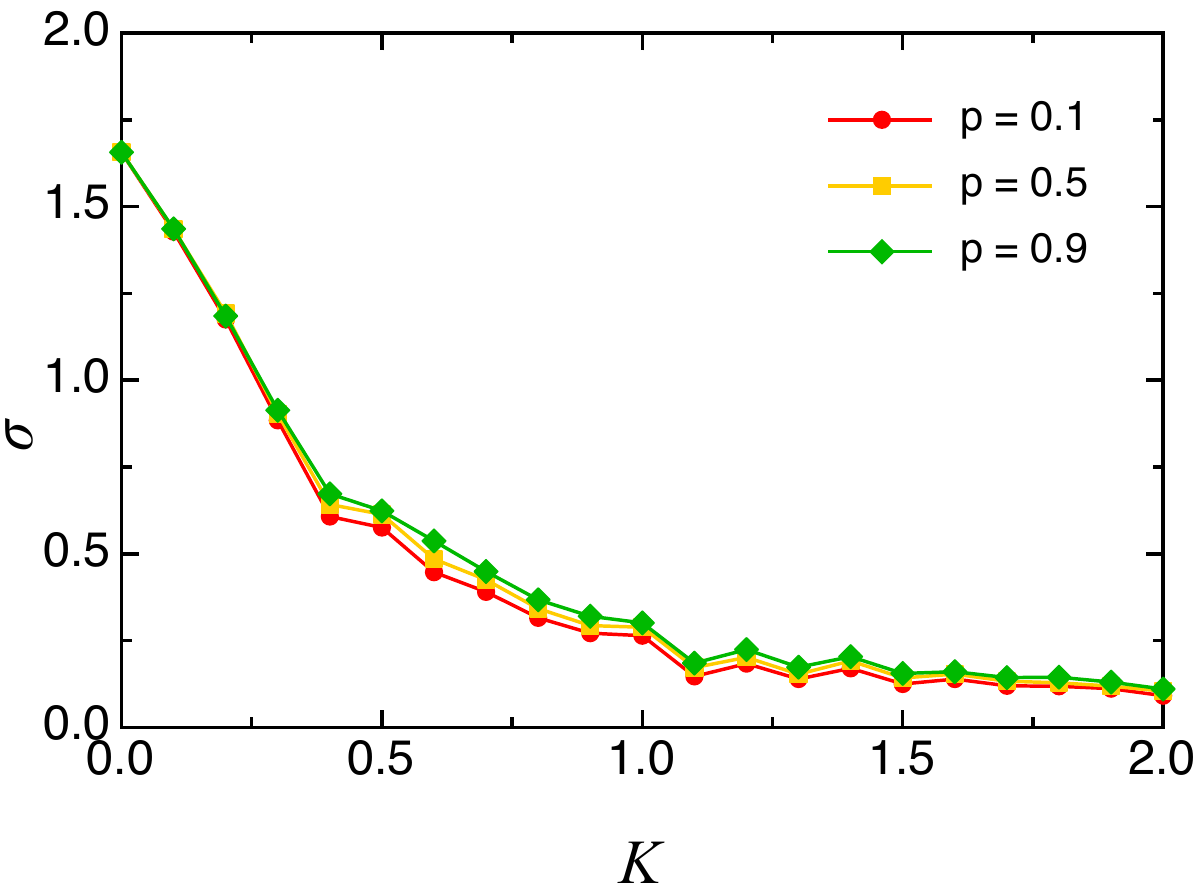}
\end{tabular}{}
\caption{ Synchronization order parameter $\sigma$ with respect to the coupling strength $K$, for small-world (WS) networks, consisting of $100$ neurons, generated with different rewiring probabilities $p$. The information time-delays $\tau_{ij}$ are drawn randomly from (a) a uniform distribution and (b) a normal distribution with variance $5$.}
\label{fig:fhn_ws_p_sp}
\end{center}
\end{figure}

{\em Synchronization Transition in Small-World Networks:} It is of relevance to find how the synchronization transition is affected by the rewiring probability $p$ in WS small world networks, with distributed delays. Fig.\ref{fig:fhn_ws_p_sp} displays the synchronization transitions in small-world networks generated with different rewiring probabilities $p$. Panel \ref{fig:fhn_ws_p_sp}(a) displays the synchronization order parameter when time-delays are drawn randomly from a uniform distribution, and panel \ref{fig:fhn_ws_p_sp}(b) shows the case of $\tau_{ij}$s drawn randomly from a normal distribution. It can again be seen that when $\tau_{ij}$s are drawn from a normal distribution, the synchronization error is marginally lower, compared to the case of uniformly distributed delays.

\begin{figure}[htb]
\begin{center}
\hspace{-0.3cm}
\includegraphics[height=6cm,width=7.5cm]{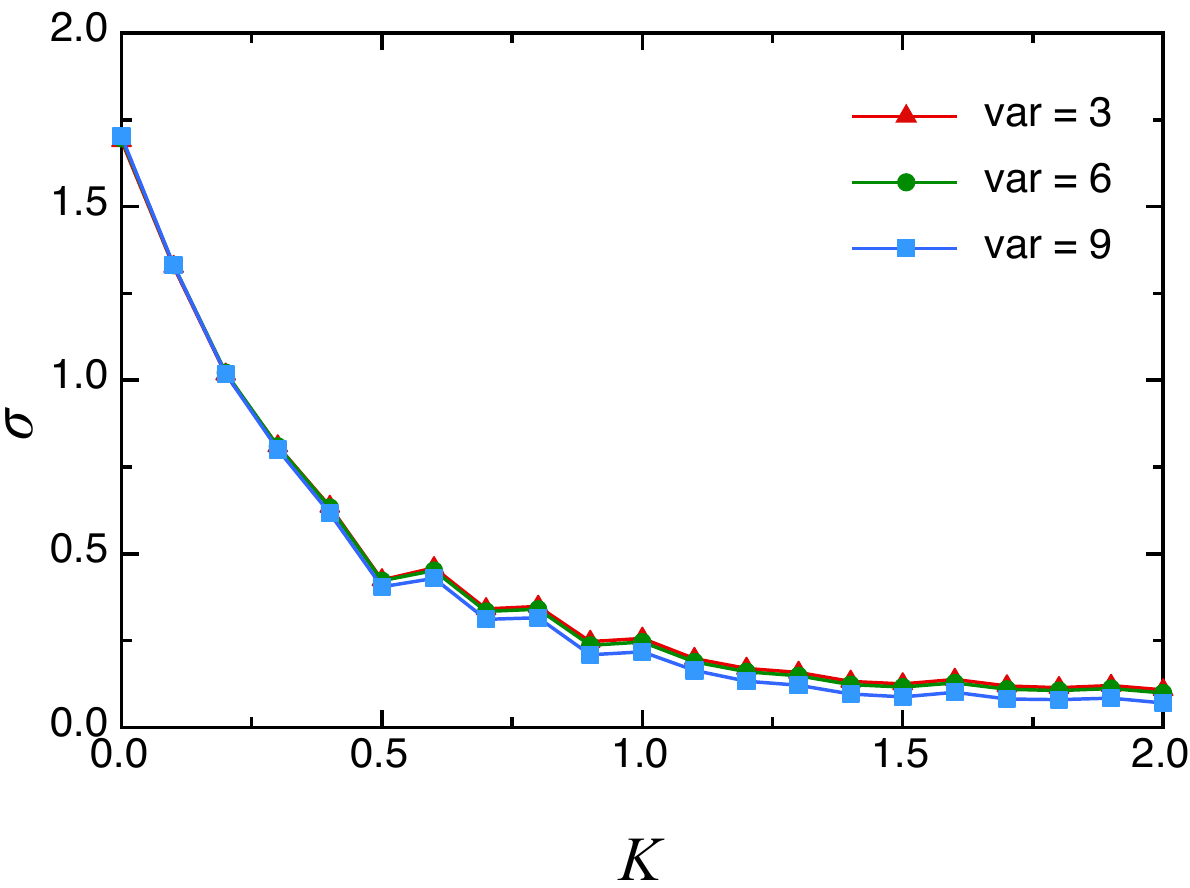}
\caption{Synchronization parameter $\sigma$ as a function of the coupling strength $K$ for scale-free (BA) networks of $N=100$ neurons. The information time-delays $\tau_{ij}$ are drawn from a normal random distributions with different values of variance.}
\label{fig:fhn_ba_norm}
\end{center}
\end{figure}

{\em Effect of the variance of the distribution of the delays on the synchronization transition:} Lastly, we study the effect of the spread of the distribution of delays, by comparing the synchronization order parameter for the case of delays drawn randomly from normal distributions with different variances or standard deviations (cf. Fig.\ref{fig:fhn_ba_norm}). From the figure one can only infer that {\em variance does not significantly affect synchronization}, with broader distributions yielding only marginally lower synchronization errors.

\section{Conclusions}

\noindent In summary, we have investigated the synchronization of networks of FitzHugh-Nagumo neurons coupled in scale-free, small-world and random topologies, in the presence of distributed time delays in the coupling of neurons. We explored how the synchronization transition is affected when the time delays in the interactions between pairs of interacting neurons are non-uniform. We find that the presence of distributed time-delays does {\em not} change the behavior of the synchronization transition significantly, vis-a-vis networks with constant time-delay, where the value of the constant time-delay is the mean of the distributed delays. We also notice that a normal distribution of delays gives rise to a transition at marginally lower coupling strengths, vis-a-vis uniformly distributed delays. These trends hold across classes of networks and for varying standard deviations of the delay distribution, indicating the generality of these results. So we conclude that distributed delays, which may be typically expected in real-world situations, do not have a notable effect on synchronization. This allows results obtained with constant delays to remain relevant even in the case of randomly distributed delays.

\begin{acknowledgments}
I wish to thank Sudeshna Sinha for her useful comments and discussions.
\end{acknowledgments}


\end{document}